\documentclass[twocolumn,prl,aps,showpacs,10pt]{revtex4-1}

\usepackage{graphicx}
\usepackage{subfigure}
\usepackage{amsmath}
\usepackage{xspace}
\usepackage{xr}
\usepackage{color}
\usepackage{upgreek}
\usepackage{ulem}
\usepackage{hyperref}

\newcommand{\FS}{FS\xspace}
\newcommand{\FSone}{FS1\xspace}
\newcommand{\FStwo}{FS2\xspace}
\newcommand{\FSs}{FSs\xspace}

\newcommand{\DoS}{DoS\xspace}
\newcommand{\QPR}{QPR\xspace}
\newcommand{\QPRone}{QPR1\xspace}
\newcommand{\QPRtwo}{QPR2\xspace}
\newcommand{\QPRs}{QPRs\xspace}

\newcommand{\SC}{SC\xspace}

\def\Ef{$E_\mathrm{F}$\xspace}
\def\didv{\mbox{$\mathrm{d}I/\mathrm{d}V$}\xspace}
\def\didvv{\mbox{$\mathrm{d}I/\mathrm{d}V(V)$}\xspace}

\def\etal{\mbox{\it et al.}\xspace}

\begin{document}
\title{Experimental demonstration of a two-band superconducting state for lead\\
			 using scanning tunneling spectroscopy}
\author{Michael Ruby$^1$,
        Benjamin W. Heinrich$^1$,
				Jose I. Pascual$^{1,2}$,
				Katharina J. Franke$^{1}$}
\affiliation{$^1$Fachbereich Physik, Freie Universit\"at Berlin,
                 Arnimallee 14, 14195 Berlin, Germany.\\
						 $^2$CIC nanoGUNE and Ikerbasque, Basque Foundation for Science,
						     Tolosa Hiribidea 78, Donostia-San Sebastian 20018, Spain}
\date{\today}
\begin{abstract}

The type I superconductor lead (Pb) has been theoretically predicted to be a two-band superconductor. We use scanning tunneling spectroscopy (STS) to resolve two superconducting gaps
with an energy difference of $150\,\mathrm{\upmu eV}$.
Tunneling into Pb(111), Pb(110) and Pb(100) crystals reveals a strong dependence of the
two coherence peak intensities on the crystal orientation.
We show that this is the result of a selective tunneling into
the two bands at the energy of the two coherence peaks.
This is further sustained by the observation of signatures of the Fermi sheets
in differential conductance maps around subsurface defects.
A modification of the density of states of the two bands by adatoms
on the surface confirms the different orbital character of each of the two sub-bands.

\end{abstract}
\maketitle 

The theory of Bardeen, Cooper and Schrieffer (BCS) has been extremely successful in describing many aspects of superconductivity (\SC). It predicts the formation of a condensate of quasi-particles, the so-called Cooper pairs, as a result of electron-phonon coupling. The corresponding quasi-particle excitation spectrum exhibits a characteristic gap of width $2\Delta$ around the Fermi level, with $\Delta$ being the order parameter reflecting the bonding strength of the Cooper pairs. 
However, soon after the development of the BCS formalism, it was realized that the theory has to be extended for describing the properties of even the simplest elemental superconductors such as Pb, V, Ta, {\it etc.} In particular, two quasiparticle resonances have been observed in planar Pb tunneling junctions \cite{TownsendPR62PbPlanarJunction,rochlin67,BlackfordPR69PbPlanarJunction, LykkenPRB71PbPlanarJunction}. The initial interpretations of these experiments proposed an anisotropic electron-phonon coupling leading to a $\textbf{k}$-dependent order parameter as the origin of this behavior~\cite{Bennett}.

With the discovery of superconductivity in highly anisotropic, composite materials, such as $\mathrm{MgB}_2$, $\mathrm{NbSe}_2$, $\mathrm{CaC}_6$, \textit{etc.} with two distinct energy gaps and unexpectedly high critical temperatures \cite{giubileo01,giubileo02,MgB2Review, NovelAnisotropicSC1discovery,CaCBCSSuperconductor, CaCPhononMediatedSuperconductivity, MgB2Anisotropy, CaC6Superconductivtiy, MartinezSamper2003233}, the importance of the concept of multi-band superconductivity, which had been proposed already in 1959~\cite{suhl59}, was realized. This motivated a renewed theoretical treatment of conventional superconductors with state-of-the-art methods. These revealed that two disjoint Fermi sheets (\FS) with different electron-phonon coupling strengths lead to two distinct energy gaps and an increased critical temperature as compared to a single isotropic gap \cite{choi_origin_2002}.

Floris \textit{et al.} identified by density functional theory (SCDFT) that two-band superconductivity also plays a role in the elemental superconductor Pb~\cite{Floris}. They found that the Fermi surface of Pb is composed of a compact Fermi sheet with mostly $s$-$p$-character and a tubular Fermi sheet of $p$-$d$-character. The different orbital nature leads to different electron-phonon coupling strengths~\cite{sklyadneva12}, and causes different pairing energies in the SC condensate.

Experimentally, it is difficult to distinguish between a two-band model and an anisotropic variation of the order parameter. 
Angle-resolved photoemission spectroscopy, a prime candidate for experimental band structure determination, lacks the required energy resolution. Planar tunneling junctions have revealed two peaks in the gap structure~\cite{TownsendPR62PbPlanarJunction,rochlin67,BlackfordPR69PbPlanarJunction, LykkenPRB71PbPlanarJunction}, but the tunneling current is the sum of all tunneling paths including step edges, vacancies, impurities, etc. This prohibits an unambiguous interpretation of the tunneling spectra.
 
Here, we overcome this shortcoming using scanning tunneling microscopy (STM) and spectroscopy (STS) to probe atomically-flat surfaces as well as well-defined defects and distinguish between the different contributions to tunneling.
We present direct evidences for the two-band nature of superconductivity in Pb. Two BCS-like resonances with an energy separation of $150\,\mathrm{\upmu eV}$ are observed.
Depending on the surface orientation the intensity of these peaks varies due to $\textbf{k}$-selective tunneling into the two Fermi sheets. 
Scattering patterns around sub-surface Ne impurities at the energies of the two coherence peaks reveal signatures of the shape of the respective Fermi sheets as a result of an anisotropic electron propagator in the crystal \cite{heil95,kurnosikov09, WeismannFermiSurface, FermiSurfaceImagingTheory}. 
Furthermore, we show that the distinct orbital character of the Fermi sheets is reflected by the modification of density of states at adatoms, which tends to increase the weight of tunneling into more localized $d$-states over the delocalized $s$-$p$-derived states.

Our experiments were carried out in a \textsc{Specs} JT-STM under ultra-high vacuum conditions at a base temperature of $1.2\,\mathrm{K}$. Pb is a type I superconductor with a critical temperature of $T_{\mathrm{c}} = 7.2\,\mathrm{K}$ and a coherence length of $83\,\mathrm{nm}$. The single crystals were cleaned by cycles of Ne$^+$ ion sputtering  at $900\,$eV with a Ne pressure of $1.5\times 10^{-4}\,$mbar (background pressure: $< 1.5\times 10^{-9}\,\mathrm{mbar}$) and annealing to $430\,\mathrm{K}$ for $30\,\mathrm{min}$ until a clean, atomically flat and superconducting surface was observed.
To achieve high energy resolution we cover etched W-tips with Pb by deep indentations into the clean Pb surface until superconductor-superconductor tunneling spectra are measured~\cite{KatharinaSC} (see Supplemental Material for details~\cite{Supplementary}). The use of a superconducting tip together with an elaborated grounding and RF-filtering scheme yields an effective energy resolution of $\approx\!45\,\mathrm{\upmu eV}$ at $1.2\,\mathrm{K}$ (compared to $\approx\!360\,\mathrm{\upmu eV}$ with a normal metal tip, in which Fermi-Dirac broadening limits the energy resolution).
Crystalline directions were determined by atomically resolved topographies of the clean Pb surface  (see Fig.\,S1~\cite{Supplementary}).

\begin{figure}[t]
	\includegraphics[width=0.998\linewidth]{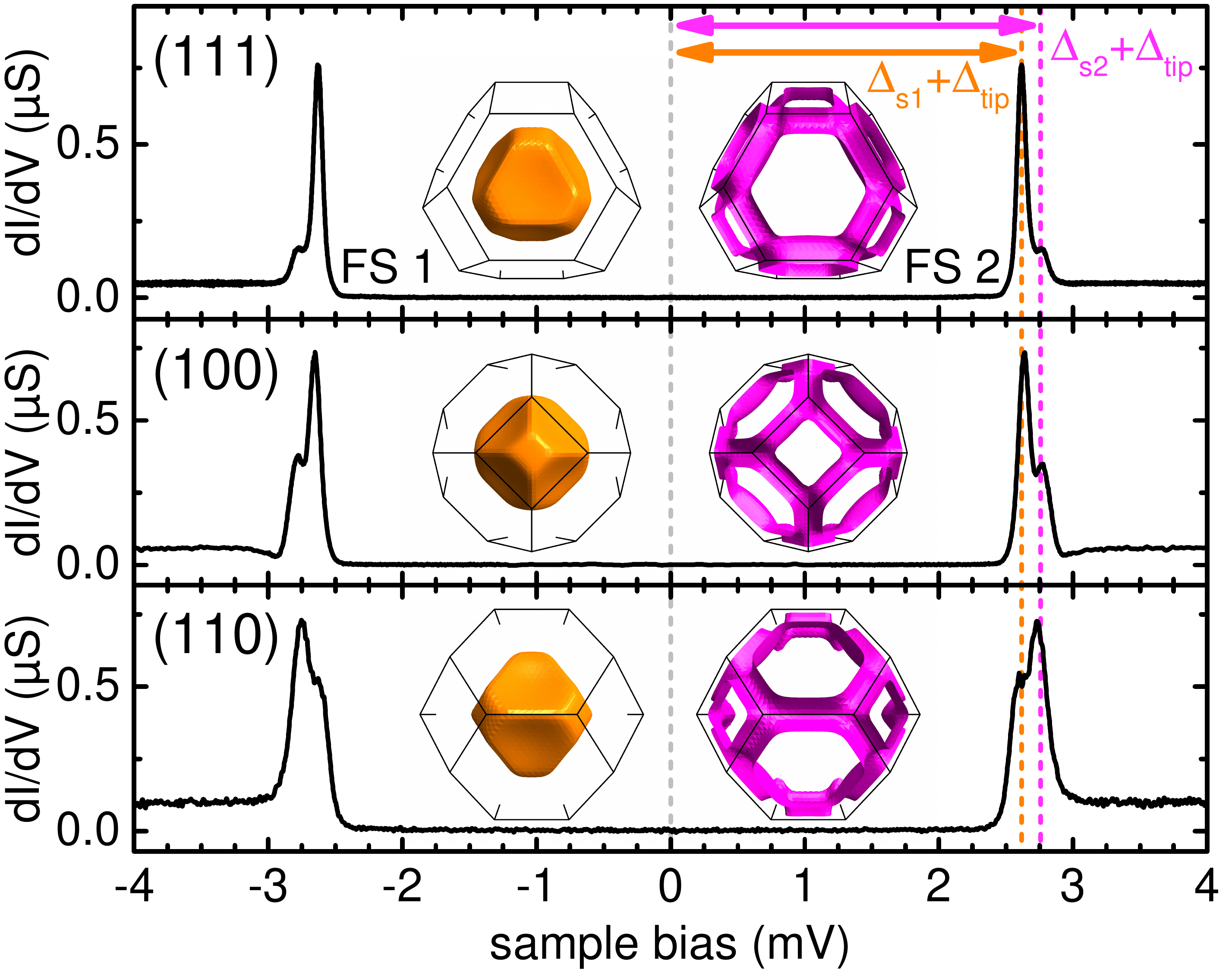}
	\caption{\didvv-spectra on clean terraces of Pb(111), (100), and (110) single crystal surfaces. 
			The superconducting gap around \Ef is framed by \QPRs
			at $\approx\!\pm2.7\,\mathrm{mV}$, consisting of two peaks separated by $\approx\!150\,\mathrm{\upmu eV}$.
			The energy of the peaks is given as the sum of the pairing energy of the tip ($\Delta_\mathrm{tip}$) and the sample ($\Delta_{\mathrm{s1}}$ and $\Delta_{\mathrm{s2}}$, respectively). The insets show the corresponding top views on the two \FS sheets of a Pb single crystal. 3D-models from \cite{FermiSurfaceDB}. Lock-in modulation amplitude was $15\,\mathrm{\upmu V}_\mathrm{rms}$ at $912\,\mathrm{Hz}$. 
					}
	\label{fig:Fig1}
\end{figure}

We record spectra of the differential conductance (\didvv) on clean terraces of the three low-index surfaces (111), (100), and (110) of Pb single crystals to probe their superconducting energy gaps (Fig.\,\,\ref{fig:Fig1}).
Around \Ef, the superconducting gap (zero conductance) is framed by quasi-particle resonances (QPR) at $\approx\!\pm 2.7\,\mathrm{meV}$ 
\footnote{Direct tunneling of Cooper pairs giving rise to a $dc$ Josephson-current is not observed due to the weak overlap of the condensate wave functions at the experimental tip--sample distance.}.
In the spectra of all surface orientations, we observe two pairs of \QPRs separated by $\approx\!150\,\mathrm{\upmu eV}$~\footnote{In the \didvv spectrum on Pb(100) a dip is observed at $\pm 3\,\mathrm{mV}$. Such dips are occasionally resolved on all three surfaces and depend on the tip employed. They are due to deviations from pure BCS-like \DoS originating from interband scattering~\cite{NoatPRB10}.}. Due to the superconducting state of the tip, the spectra are a convolution of two \SC density of states. In particular the position of the \QPRs is shifted by $\Delta_\mathrm{tip}$. To extract the exact energy positions and intensities of the two \QPRs, we deconvolute the spectra as described in the Supplemental Material~\cite{Supplementary}. We can unambiguously link the appearance of the two peaks with $\approx\!150\,\mathrm{\upmu eV}$ separation to a property of all samples, independent of the tip's single gap.
Similar splittings have been observed earlier in planar Pb tunnel junctions and 
attributed to the anisotropy of the \FS and of the electron‐phonon coupling of Pb~\cite{
LykkenPRB71PbPlanarJunction,
BlackfordPR69PbPlanarJunction,
TownsendPR62PbPlanarJunction,
Bennett}.
More recently however, Floris \etal predicted that Pb is a two-band superconductor with two well-separated \FSs, with one of them being highly anisotropic and the other almost spherical (see inset in Fig.\,\,\ref{fig:Fig1})~\cite{Floris}.
The inner \FS (\FSone) has an almost spherical shape.
The outer \FS (\FStwo) has a tubular shape~\cite{FermiSurfaceAndersonGold}. \FSone is mostly of $s$-$p$ character with a smaller pairing energy than \FStwo which is of $p$-$d$-like character~\cite{Floris}.
A manifestation of the different pairing energy associated to each \FS is the different position of the corresponding quasi-particle resonances \QPRone and \QPRtwo in the \didvv spectra. Hence, we identify the inner and outer peaks as tunneling into \FSone and \FStwo of the sample, respectively. The existence of a single gap in the STM tip is in agreement with its expected micro-crystalline character~\cite{DirtySuperconductors}. While the energy separation between \QPRone and \QPRtwo is constant for all surfaces, we observe distinct relative peak intensities for the different surface orientations (Fig.\,\,\ref{fig:Fig1}). The tunneling probability depends on transition matrix elements, which depend on the $k_\perp$ component of the wavevector $\boldsymbol{k}$. A strong tunneling contribution thus requires access to the \FS sheets with the wavevector $\boldsymbol{k}$ being mostly perpendicular to the surface. The insets in Fig.\,\,\ref{fig:Fig1} show the top views of the two \FS sheets for the given crystal orientations. 
\FSone is compact, which implies that tunneling with strong $k_\perp$ contribution into the (111)-, (100)- and (110)-surface is possible. In contrast to this, \FStwo exhibits open pores along the $\boldsymbol{k}_{\Gamma\rightarrow L}$- and $\boldsymbol{k}_{\Gamma\rightarrow X}$-direction. Hence, for these directions tunneling into \FStwo is only possible with wavevectors with considerable $k_\parallel$ component, which is accompanied by a reduced tunneling probability. 
Therefore, the ratio of intensities of \QPRone and \QPRtwo is largest on the (111) surface,
where \FStwo exhibits the largest pore, followed by the (100)-surface.  
On the (110)-surface, \QPRtwo is even more intense than \QPRone (see Supplemental Material for a quantitative analysis of the intensities~\cite{Supplementary}). Both \FSs can be accessed by electrons with $\boldsymbol{k}$ vectors with mostly $k_\perp$ contribution and therefore participate almost equivalently in tunneling.

\begin{figure*}[tb]
	\includegraphics[width=0.99\linewidth]{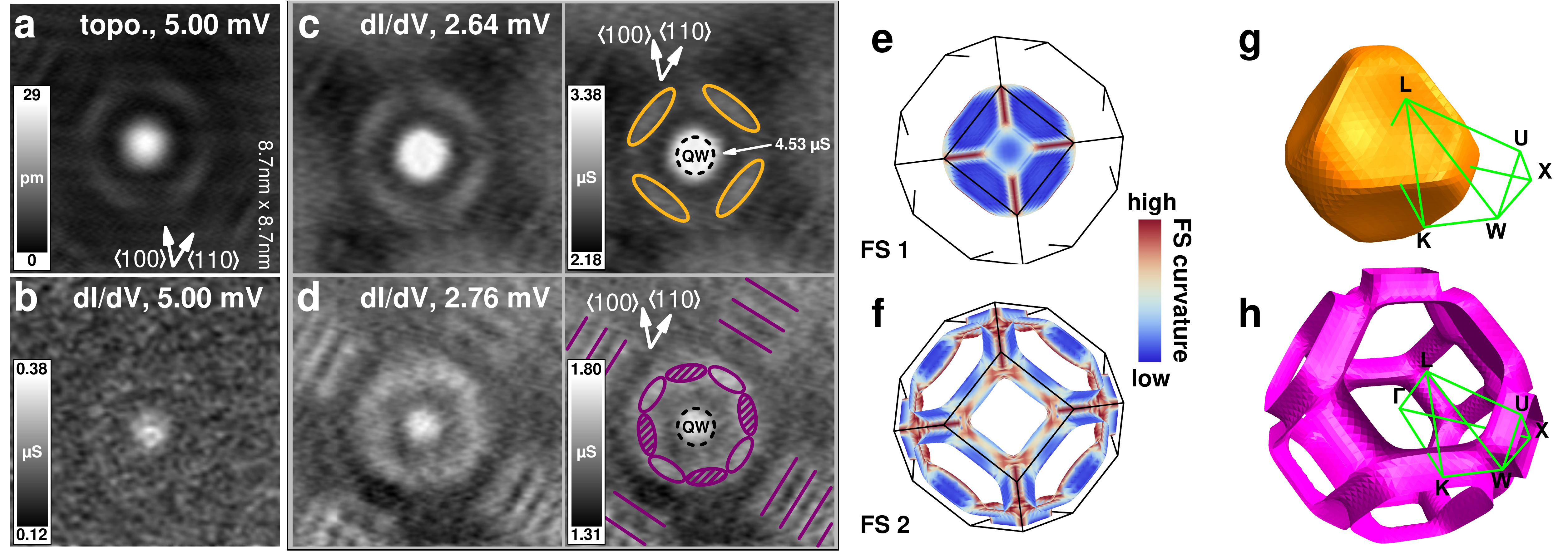}
	\caption{{(a--d)} Subsurface Ne inclusion.
					High-resolution topography (a) on an atomically flat Pb(100) terrace
					(setpoint: $5\,\mathrm{mV}, 250\,\mathrm{pA}$). 
					Constant-height \didv maps at $5\,\mathrm{mV}$ (b), 
					$2.64\,\mathrm{mV}$ (c), and $2.76\,\mathrm{mV}$ (d) 
					($25\,\mathrm{\upmu V}_\mathrm{rms}$, $912\,\mathrm{Hz}$, setpoint: $5\,\mathrm{mV}, 500\,\mathrm{pA}$).
					In the duplicated maps, prominent scattering signatures are highlighted as guide for the eye.
					A quantum well state between the impurity and the surface
					leads to the higher conductance in the center of all maps. 
					Note that the topography in (a) combines features visible in (c) and (d) because all states up to the applied bias---i.e., both Fermi sheets---contribute to the tunneling current.
					(e--h) 3D-models of the two \FSs of Pb from Ref.\,\,\cite{FermiSurfaceDB}.
					The curvature of the \FS is color-coded onto the 3D model in (e) and (f),
					which are oriented according to the crystalline directions in (a--d). 
					\textit{Dark blue} and \textit{dark red} correspond
					to low and high curvature, respectively.
					Black lines in (e) and (f) mark the boundaries of the first Brillouin zone.
					}
	\label{fig:Fig2}
\end{figure*}

The observed energy splitting in STS together with the dependence of the \QPR intensities on the surface orientation agrees well with the two-band superconductivity connected to the \FSs of Pb as predicted by Floris \etal~\cite{Floris}. The difference between the pairing energies of $\approx 10\,\%$ is however smaller than calculated ($\approx 30\,\%$~\cite{Floris}). We link this to interband scattering, which diminishes the difference in the pairing energy of the two bands~\cite{NoatPRB10}. 

We now search for a more direct evidence of the presence of the two \FSs with different order parameter. One way to image the different symmetries of the two \FSs is to inspect \didv maps around buried impurities which show characteristic modulations of the density of states \DoS around them (Fig.~\ref{fig:Fig2} and Fig.~S3 in the Supplemental Material~\cite{Supplementary}).
On the Pb(100) surface we find typical patterns in STM topographies (see Fig.\,S2 \cite{Supplementary}). These consist of a bright or dark center, which is framed by patterns of fourfold symmetry. 

The impurities are most likely Ne-filled subsurface nanocavities, which are residuals of the Ne$^+$ ion sputtering \cite{QWState, kurnosikov09}. They act as scattering centers and give rise to quantum well states between the surface and the impurity~\cite{QWState}. The subsurface inclusion appears then as a protrusion or a depression, depending on whether or not the sample bias
matches the quantization condition of the quantum well state. The quantum-well state typically has a width of several hundreds of $\mathrm{meV}$ ({\it e.g.} Fig.\,S3h in the Supplemental Material). Therefore it is present in a wide energy range.
Laterally away from the impurity center the two constant-height \didv maps at the energy of the two \QPRs show quite different patterns of charge density oscillations. The map at $2.64\,\mathrm{meV}$, which results from tunneling into \QPRone, exhibits a square-like pattern with the edges along the $\left\langle110\right\rangle$ directions around the bright center (highlighted by \textit{orange ellipses} in Fig.\,\,\ref{fig:Fig2}c).
The map at the energy of \QPRtwo ($2.76\,\mathrm{meV}$, Fig.\,\,\ref{fig:Fig2}d) shows areas of high intensity along the $\left\langle100\right\rangle$ (\textit{shaded purple ellipses}) and $\left\langle110\right\rangle$ (\textit{open purple ellipses}) directions, respectively. Additionally, long-ranging oscillations appear in the \didv signal along the $\left\langle110\right\rangle$ directions (indicated by \textit{purple stripes}).

According to Weismann and co-workers, the charge density oscillations result from scattering and focusing of bulk electrons (holes) at subsurface impurities with an anisotropic electron (hole) propagation~\cite{WeismannFermiSurface}. In analogy to Huygens principle, the group velocity of the electrons (holes) $\mathrm{d}E/\mathrm{d}k$ is perpendicular to the \FS and therefore nearly parallel for beams arising from areas of low \FS curvature. This leads to a focusing of the electron propagator into the normal direction of these regions~\cite{heil95}. Hence, the real space distribution of the \DoS on the surface above the impurity, which is resolved via \didv mapping, is directly related to the shape of the \FS.

We can now assign the features of the \didv maps to low-curvature regions of \FSone and \FStwo, respectively. The curvature of the respective \FSs is color-coded onto the 3D models in Fig.\,\,\ref{fig:Fig2}(e,f). 
\FSone contains two groups of low-curvature regions (\textit{dark blue} in Fig.~\ref{fig:Fig2}e): eight large regions with the group velocity ({\it i.e.} \FS surface normal) pointing into the $\left\langle111\right\rangle$ directions, $i.e.$, along the $\Gamma$-$L$-direction, and six smaller regions with the surface normal pointing towards $\left\langle100\right\rangle$, $i.e.$ along the $\Gamma$-$X$-direction.
Projected on the (100) surface, we can assign the four square-like stripes of high conductance ({\it orange ovals in Fig.~\ref{fig:Fig2}c)} to the focusing of electron propagation in the $\left\langle111\right\rangle$ directions. Also for the $\left\langle100\right\rangle$ direction a higher intensity is expected. However, this direction is dominated by the quantum well state as discussed above. 
Both signals are superimposed and thus not distinguishable. 

\FStwo can be described as a complex structure with tubes connecting the $U$- and $W$-points, and connecting $K$- and $W$-points, respectively. Despite having a large three-dimensional curvature (Fig.\,\,\ref{fig:Fig2}h), the tubes exhibit one dimensional low curvature regions (lines). Along the line the group velocity is therefore pointing in the same direction, giving rise to an enhanced scattering pattern, in analogy to a decreased decay of Friedel oscillations in two dimensions \cite{FermiSurfaceImagingTheory}. The fourfold symmetry of the $U$-$W$- and $K$-$W$-tubes, respectively, then gives rise to the octagonal pattern ({\it purple ovals} in Fig.\,\,\ref{fig:Fig2}d) projected on the surface as seen in the \didv map. 
The oscillations at larger distance can be linked to long-range interference of electrons (holes), due to scattering at side facets of the inclusion~\cite{kurnosikov08,kurnosikov09} or
due to interference of electrons (holes) with the same group
velocity, but originating from different areas on the FS \cite{FermiSurfaceImagingTheory}.
The distinct appearance of these patterns at \QPRone and \QPRtwo are thus a further proof of the geometrically very different \FSs with different pairing energy.

\begin{figure}[t]
	\includegraphics[width=0.99\linewidth]{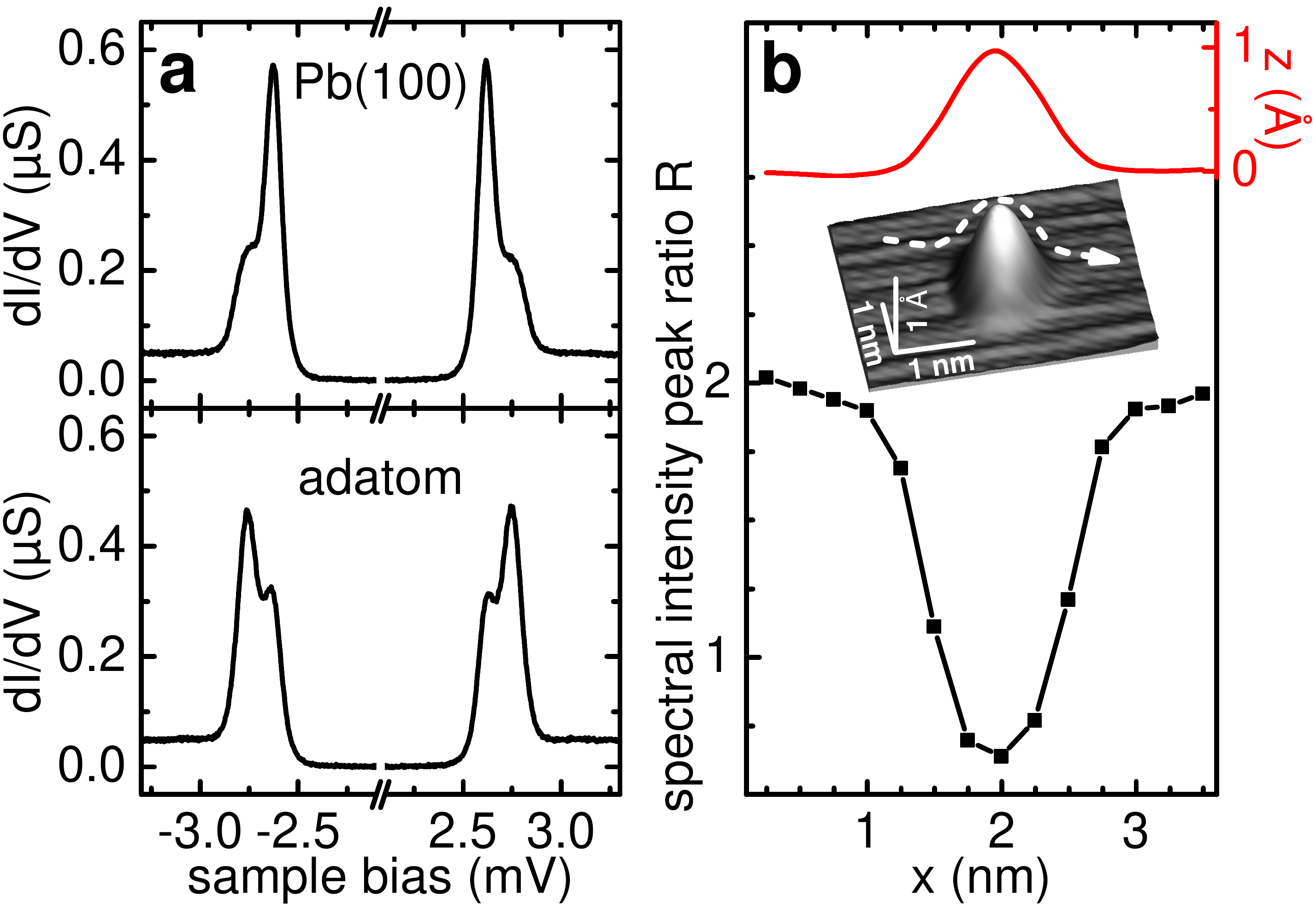}
	\caption{Pb adatom on Pb(100).
					(a) \didvv spectrum on clean Pb(100) (top) and on a Pb adatom (bottom).
					(b) Intensity ratio $R$ of \QPRone vs. \QPRtwo and apparent height across
					a Pb adatom as sketched in the inset.
					$R$ is determined by numerical deconvolution of the
					spectra as discussed in the Supplemental Material~\cite{Supplementary}. 
					}
	\label{fig:Fig3}
\end{figure}

According to the SCDFT calculations in Ref. \cite{Floris} each \FS has a different orbital character. We expect that this character will be reflected in the interaction with local potentials. Adsorbates interact with the electronic bands of the surface and locally modify the corresponding density of states. To probe this interaction with both the \mbox{$s$-$p$-,} and $p$-$d$-derived \FSs, we deposited Pb adatoms from the lead-covered tip onto the surface by applying voltage pulses of $6$ to $10\,\mathrm{V}$ at a tip--sample distance of approximately $1\,\mathrm{nm}$. 
The inset of Fig.\,\,\ref{fig:Fig3}b shows a topography of an as-deposited adatom. 
The excitation spectrum above the center of the adatom (Fig.\,\,\ref{fig:Fig3}a, bottom) shows that \QPRtwo is more intense than \QPRone, in contrast to the spectrum on the clean surface (Fig.\,\,\ref{fig:Fig3}a, top). 

The spatial extension of this intensity variation is reflected in a series of spectra taken along a line over the adatom (see inset of Fig.\,\,\ref{fig:Fig3}b). The relative intensities $R = A_1/A_2$ of the \QPRs---with $A_1$ and $A_2$ being the intensity of \QPRone and \QPRtwo, respectively, after deconvolution---are shown in Fig.\,\,\ref{fig:Fig3}b. The value decreases continuously from $2$ over the clean surface to $0.65$ on top of the adatom.
Thus the adatoms increase the tunneling probability into \FStwo compared to \FSone.
The variation of $R$ is constricted to the adatom site \footnote{Atomic modulation of the gap structure of multiband superconductors has been observed earlier~\cite{guillamon08}. Yet the modulation strength of the effect observed here is unexpectedly high and hints towards a strong modification of the local density of states at the adatom site.}.
In principle, a protruding feature such as an adatom geometrically favors tunneling with large $k_\perp$ contribution and may thus enhance the tunneling probability into \FSone, leading to a larger $R$. The opposite trend observed in the experiment suggests another scenario.
The increased intensity of \QPRtwo indicates that tunneling into the band of \FStwo is particularly enhanced at the localized potential of the Pb adatom. This is a consequence of a strong confinement of localized $d$-derived states around an impurity potential. Therefore the band that is associated to \FStwo and is hybridized with $d$-states is more affected than the extended $s$-$p$-band that creates \FSone~\cite{sklyadneva12}.

Despite Pb being one of the best characterized type~I superconductors, the theoretical prediction of Pb being a two-band superconductor was experimentally not unambiguously evidenced up to date. The early-on observed splitting of \QPRs could either be described by an anisotropic electron-phonon coupling term or two distinct electronic bands at the Fermi level. We have shown clear fingerprints of the two-band nature of superconductivity in Pb. STS resolved the differing pairing energy on the two bands as $150\,\mathrm{\upmu eV}$, which is smaller than theoretically predicted. Calculations of interband scattering events may be able to explain this deviation. The energetically separated \FSs allowed for a direct mapping of their symmetry in real space. This method is complementary to quasiparticle interference mapping by STM~\cite{McElroy03}, which is frequently used to resolve characteristics of the \FS of, \textit{e.g.}, high-T$_c$ superconductors. Most importantly, the intensity of quasiparticle interference falls off too rapidly in three-dimensional electron systems (such as Pb), and, therefore, requires two-dimensional states. Furthermore, it does not involve reconstruction of the Fermi surface by Fourier transformation, but directly reflects the symmetry of reciprocal space in real space.

The tuning of orbital contributions around atoms allows us in a proof-of-principle experiment to favor tunneling into one or the other \FS and might be used---together with the focusing properties of the curved \FSs---for $\textbf{k}$-selective filtering in future tunneling devices.

We gratefully acknowledge funding by the DFG through Sfb 658 and grant FR-2726/4.

\bibliographystyle{apsrev4-1}
\bibliography{paper}



%
\end{document}